\documentclass[12pt]{article}
\usepackage[dvips]{graphicx}

\newcommand{\be}{\begin{equation}}

\newcommand{\ee}{\end{equation}}

\newcommand{\eq}{\begin{equation}}
\newcommand{\eqx}{\end{equation}}
\newcommand{\eqn}{\begin{eqnarray}}
\newcommand{\eqnx}{\end{eqnarray}}
\newcommand{\f}[2]{\frac{#1}{#2}}
\newcommand{\eps}{\varepsilon}
\newcommand{\tr}{\mbox{\rm tr}\,}

\newcommand{\cor}[1]{\left\langle {#1} \right\rangle}
\renewcommand{\th}{\theta}

\newcommand{\rr}[4]{#1, {\it #2 \/}{\bf #3} #4}

\title{Multiplying unitary random matrices\\ -- universality and spectral
  properties} 

\author{Romuald A. Janik\footnote{e-mail: {\tt
ufrjanik@if.uj.edu.pl}}  \ and Waldemar Wieczorek\footnote{e-mail: {\tt
wieczor@th.if.uj.edu.pl}}\\
M. Smoluchowski Institute  of Physics,\\ 
Jagellonian University,\\ 
Reymonta 4,\\
30-059 Krak\'{o}w, Poland
}

\begin{document}

\maketitle

\begin{abstract}
In this paper we calculate, in the large $N$ limit, the eigenvalue
density of an infinite product of random unitary matrices, each of
them generated by a random hermitian matrix. This is equivalent
to solving unitary diffusion generated by a hamiltonian random in
time. We find that the result is universal and depends only on the
second moment of the generator of the stochastic evolution. We find
indications of critical behavior (eigenvalue spacing scaling like
$1/N^{3/4}$) close to $\th=\pi$ for a specific critical evolution time
$t_c$.    
\end{abstract}

\section{Introduction}

A key feature of random matrix theory is that many properties of
random matrix models do not depend on the fine details of these models
but only on some very general symmetry properties and a very limited
number of numerical coefficients (usually just a single coefficient is
enough). These universal properties facilitate the widespread
applications of random matrix models in various fields since one can
use the models to learn something about the behavior of complex
systems without knowing all the precise microscopic details of these
systems. 

The simplest example of such a behavior is the eigenvalue density of
a matrix with entries independently distributed according to some
probability distribution. Then the eigenvalue density in the $N \to
\infty$ limit follows Wigner's semicircle law with the scale set just
by the second moment of the distribution. All the dependence on other
properties of the initial probability distribution disappears. In this
work we will uncover a similar type of universality in a different
context.

The aim of this paper is study infinite products of
unitary random matrices, and derive their properties in the large $N$
limit. This can be interpreted as a multiplicative diffusion process
on the unitary group\footnote{Recently matrix valued multiplicative
diffusion has been considered for $2\times 2$ real matrices in
\cite{JACKSON} and for infinite hermitian and complex matrices in
\cite{man}.}. A natural physical interpretation would be of a 
quantum-mechanical evolution governed by a hamiltonian which changes
randomly in time. Another
possibility would be the modelling of Wilson loops in lattice gauge
theory. 
In this paper we will not examine further these possible application
but rather concentrate on mathematically solving the model.

We show that the eigenvalue density exhibits universality properties
i.e. it only depends on the second moment of the random hamiltonian
which generates the stochastic evolution. But of course the resulting
eigenvalue density is much more complex than the semicircle law.
We derive equations for the
eigenvalue density, give explicit expression for the lowest moments
and study the properties of the model close to a phase transition
where the eigenvalue support begins to cover the whole unit circle.

\section{Multiplicative unitary diffusion}

We consider the product of $M$ unitary $N\times N$ random matrices
$U_{k}$ in the $M,N \rightarrow \infty$ limit: 
\be
\label{e.uprod}
U = \lim_{M\rightarrow \infty} \lim_{N\rightarrow \infty}
\prod_{k=1}^{M} U_{k} \; , 
\ee
where the $U_{k}$'s are generated by
\eq
U_{k} =  e^{i \varepsilon\, H_{k}},
\eqx
and where $\varepsilon = \sqrt{t/M}$. Such a scaling is standard for
diffusive processes and works also very well for matrix-valued
diffusion processes studied in \cite{JACKSON} and \cite{man}.
$t$ is then a real parameter corresponding to `diffusive' evolution
time and the continuum limit $M \to \infty$ exists.
The generators of the evolution $H_k$ are $N\times N$ hermitian
matrices drawn from a probability distribution
\eq
\label{e.hdist}
P(H) \sim e^{-N \tr V(H)} 
\eqx
where we assume that the first moment $m_1=\cor{\f{1}{N} \tr H}$
vanishes $m_1=0$.  
We will show below that the spectral properties of (\ref{e.uprod})
depend only on the second moment $m_2$ of the distribution
(\ref{e.hdist}) 
\eq
m_2 =\cor{\f{1}{N} \tr H^2 } \, .
\eqx
The main aim of this paper is to find the eigenvalue distribution of
the product (\ref{e.uprod})
\eq
\rho(\th,t)= \cor{ \f{1}{N} \sum_{j=1}^N i e^{i\th} \delta (e^{i \theta} - e^{i
    \theta_{j}}) } \; , 
\eqx
where the $e^{i\th_j}$ are the eigenvalues of $U$ defined through
(\ref{e.uprod}).  

In the next section we will use free random variable methods to derive
an equation from which one can get $\rho(\th,t)$. Sometimes we will
omit the second argument but of course the dependence on $t$ will be
there.

\section{The S-transform method}

The main difficulty encountered in \cite{man} when considering
products of random matrices was the necessity to deal with
nonhermitian matrices and eigenvalues covering two dimensional regions
of the complex plane. Here fortunately, since the product matrices are
always unitary, the eigenvalues lie on the unit circle and hence they
can be uniquely reconstructed from just the knowledge of the
moments. Therefore all information is encoded in the asymptotic
expansion of the Green's function
\eq
\label{e.green}
G(z)=\int_0^{2\pi} \f{\rho(\th)}{z-e^{i\th}} d\th \, . 
\eqx

Our aim now is to obtain the spectral density $\rho_{PROD}(\theta)$ of
product $U=\prod_{k=1}^M U_k$, equivalently the corresponding Green's
function $G_{PROD}(z)$, from the spectral density $\rho_{H}(\theta)$
for the generator $H$. 

To do that we use $S$-transforms introduced in
\cite{VOICULESCU}. Firstly we define an auxiliary function
$\chi(z)$ through: 
\be
\label{chi}
\f{1}{\chi} G\left(\f{1}{\chi}\right) -1 = z \, ;
\ee  
Then the $S$-transform is:
\be 
\label{s}
S(z) = \f{1+z}{z} \chi(z) \, .
\ee
Their main property is that the $S$-transform of a product of random
matrices is a product of $S$-transforms of the individual
factors. Before we apply this setup to (\ref{e.uprod}) let us note
that putting together the two previous equations we arrive at a
functional relation satisfied by $S$ and $G$: 
\eq
\label{e.seq}
\frac{1}{z S} \,  \;  G \left( \frac{1+z}{z} \,  \frac{1}{S}
\right) = 1 \, .
\eqx

In our case all single matrix Green's functions are the same since
they come from the same distribution so we can write: 
\be 
\label{e.sprod} 
S_{PROD} (z) = \lim_{M \rightarrow \infty} \prod_{i=1}^{M}
S_{i} (z) = \lim_{M \rightarrow \infty}  \left( S_{1} (z) \right)^{M} , 
\ee 

Let us now find $S_1(z)$. The eigenvalue density of $U_1=e^{i\eps H}$
is given by 
\be
\label{e.rhoone}
\rho_{1} (\theta, \varepsilon) = \frac{1}{\varepsilon}\; \rho_{H}
\left( \frac{\theta}{\varepsilon} \right)\, . 
\ee 
where $\eps =\sqrt{t/M}$. Inserting (\ref{e.rhoone}) and the definition
(\ref{e.green}) into (\ref{e.seq}) leads to an equation for $S_1$:
\be
\int \frac{1}{\varepsilon}\; \frac{ \rho_{H} \left(
    \frac{\theta}{\varepsilon} \right)}{1+z - e^{i \theta} z S_{1}} \;
\;d \theta \;  = 1. 
\ee
From the form of (\ref{e.sprod}) we see that we need to calculate
$S_1$ only to the order ${\cal O}(\eps^2)$.  
Substituting $u = \theta/\varepsilon$ and expanding in Taylor series
in $\varepsilon$ we obtain: 
\be
\int \rho_H (u) \left( 1 + i \varepsilon u z + (-u^2 z^2 - \frac{1}{2}
  u^2 z + s z) \varepsilon^2  + {\cal O} (\varepsilon^{3}) \right) du
= 1 \; , 
\ee 
where $s=s(z)$ comes from Taylor expansion\footnote{Here we used the
  assumption that the first moment of $H$ vanishes.}
$S_{1}(z) = 1 + s(z) \varepsilon^2 + {\cal O} (\varepsilon^{3})$.

We may now calculate $s(z)$
\be
s(z) = (z+1/2) \cor{u^2} \equiv (z+1/2)m_2 \; .
\ee
From (\ref{e.sprod}) we may now obtain the $S$-transform for product:
\be
\label{e.sfin}
S_{PROD}  =  \lim_{M \rightarrow \infty}  \left( S_{1}  \right)^{M} =
\lim_{M \rightarrow \infty} \left(1 + \frac{t}{M} s(z) \right)^{M} = \;
\; e^{t (z + \frac{1}{2}) m_2} \; . 
\ee 

This result shows that $S_{PROD}$ depends {\em only} on the second
moment $m_2$ of $H$. So in the limit $M\rightarrow \infty$ we obtain
universal behavior of the system independent of the spectral density
of generator of the stochastic evolution $H$, as long as the first
moment vanishes (no drift) and the second moment is finite. It will be
interesting to consider cases where these assumptions are violated
which would lead to anomalous diffusion. We leave these problems for
future investigation.  

The final step is to come back from $S_{PROD}$ to the Green's
function $G_{PROD}(z,t)$ (from now on we drop the subscript). It is
convenient to introduce auxiliary function $f(z,t)$ as:  
\be
\label{e.fdef}
G(z,t) = \frac{1+ f(t,z)}{z} \, .  
\ee
It is easy to check that $f$ fulfills an equation:
$f\left(\frac{1}{\chi(z)} \right) = z$. This means that $f$ and
$1/\chi$ are functional inverses of each other so the following relation
$1/\chi(f) = z$ is also true. This observation, together with the
result (\ref{e.sfin}) and the definition (\ref{s}) leads us to the
final equation: 
\be
\label{e.wynik}
z f = \left(1 + f\right) e^ { - t (f + \frac{1}{2}) m_2} \, . 
\ee  

This equation encodes all the spectral properties of the unitary
diffusion process (\ref{e.uprod}). In the next section we will proceed
to investigate some of its properties.

\section{The dynamical properties of the unitary diffusion}

\begin{figure}[t!]
\begin{center}
\includegraphics[width=4.3cm]{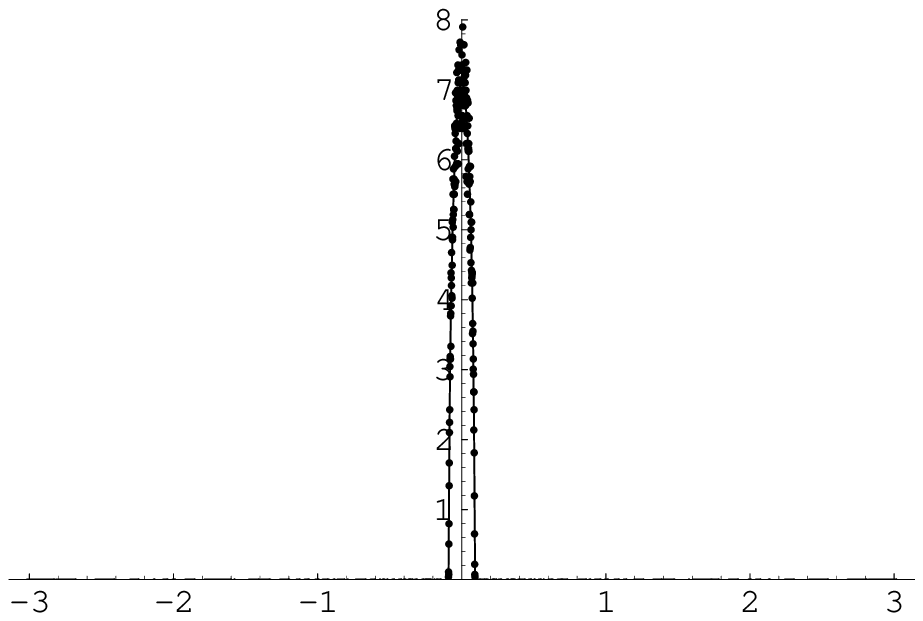}
\includegraphics[width=4.3cm]{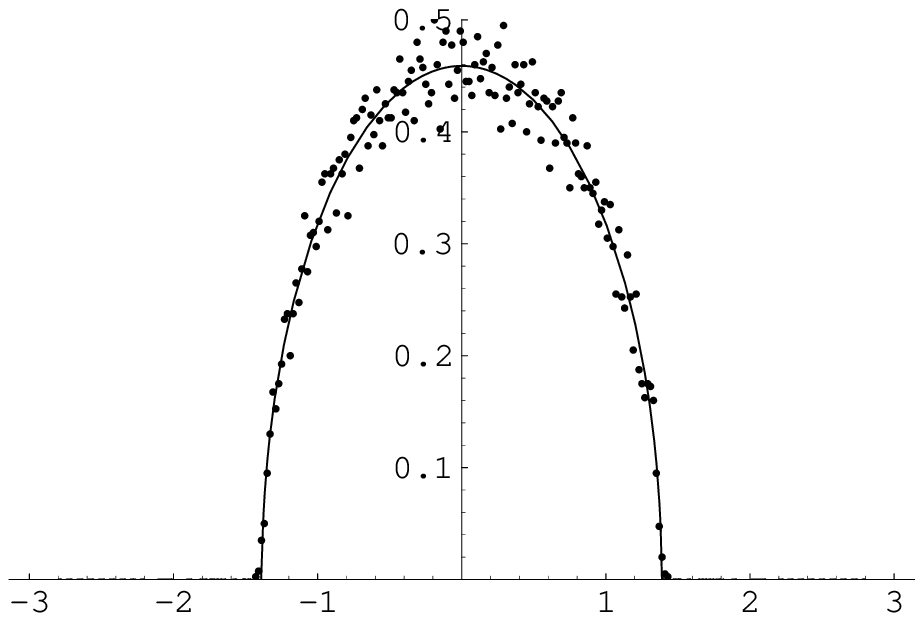}
\includegraphics[width=4.3cm]{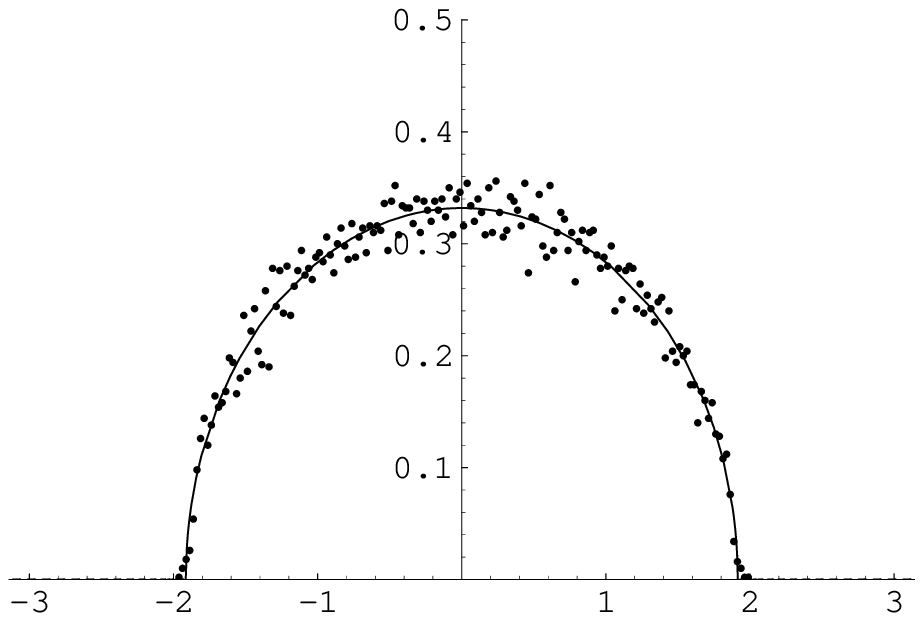}
\includegraphics[width=4.3cm]{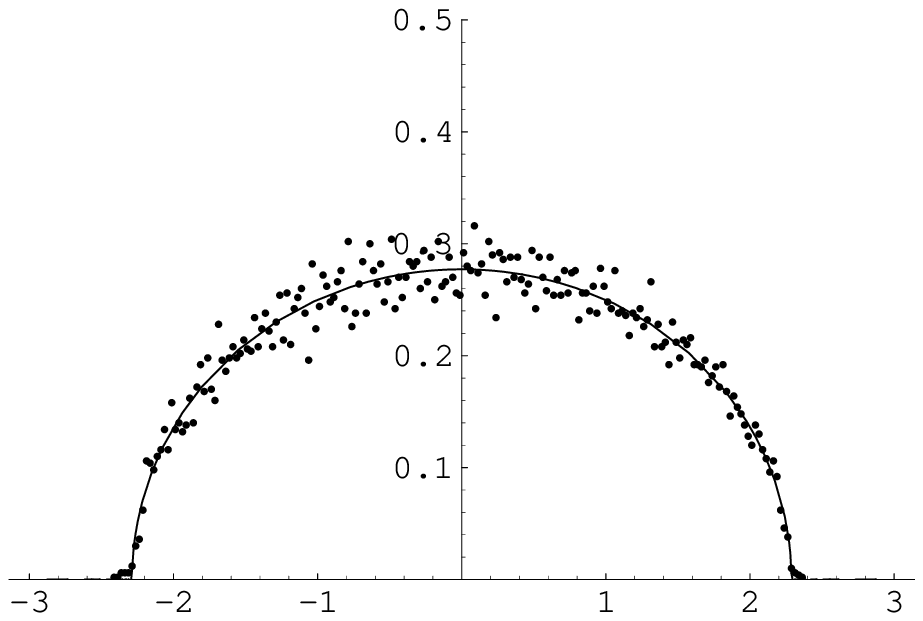}
\includegraphics[width=4.3cm]{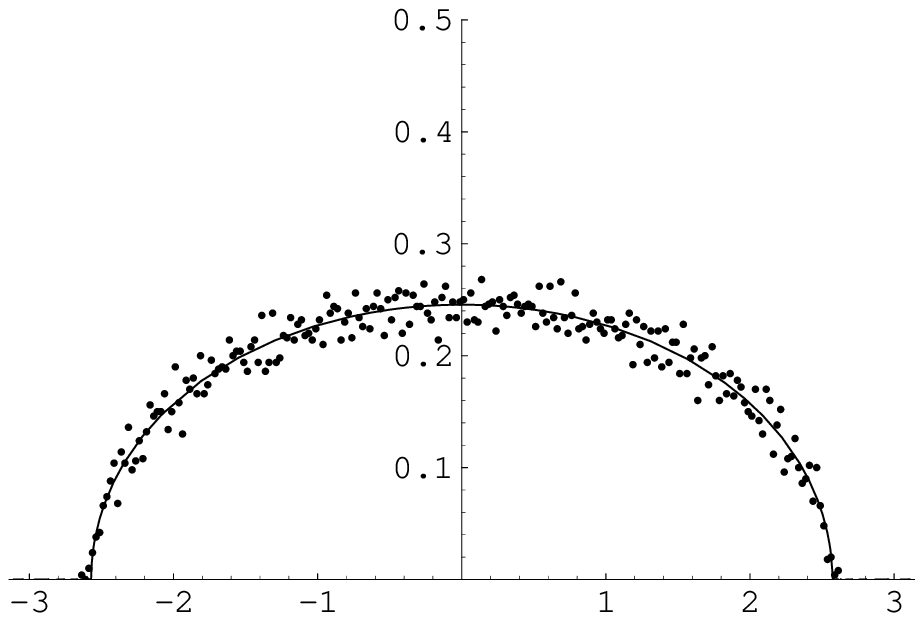}
\includegraphics[width=4.3cm]{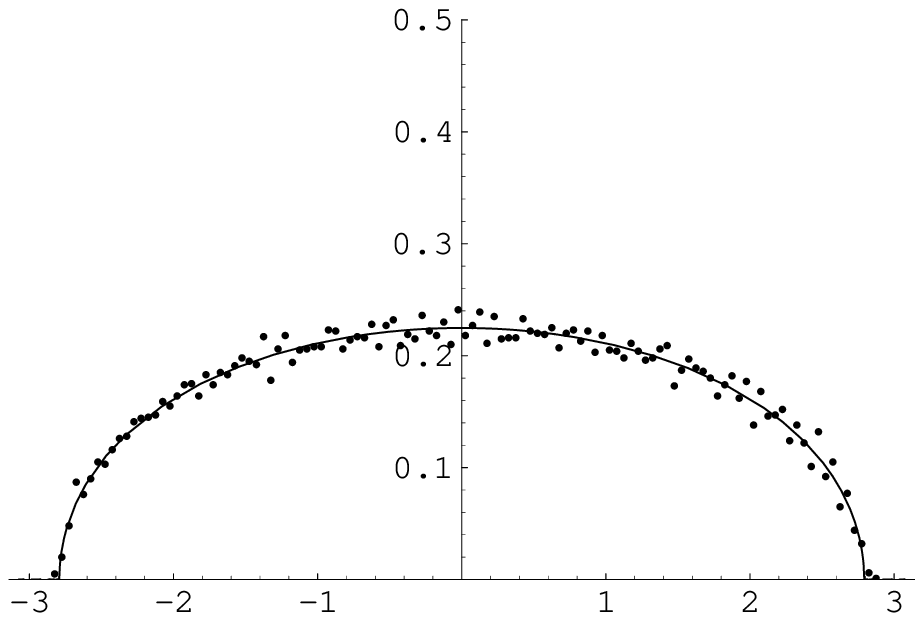}
\includegraphics[width=4.3cm]{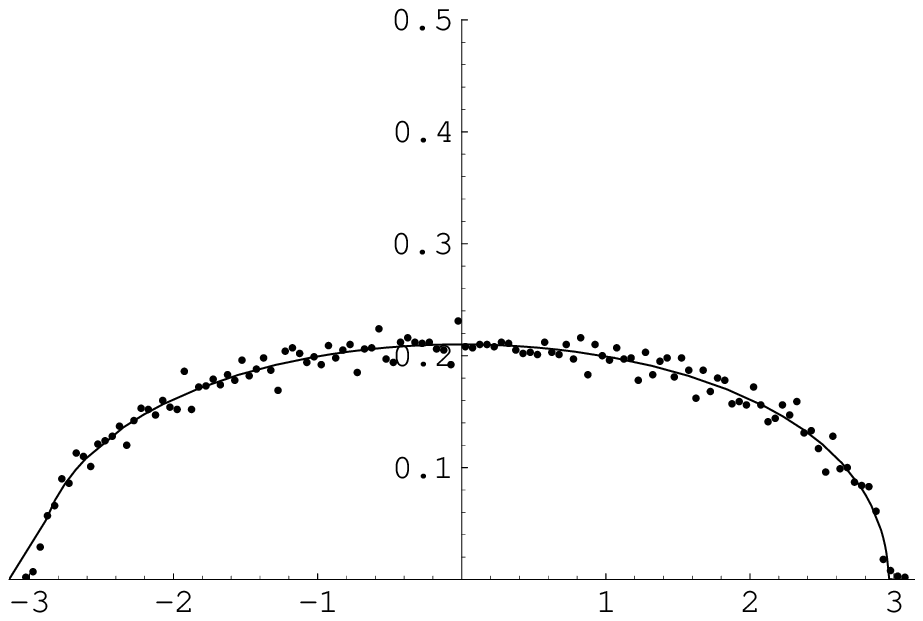}
\includegraphics[width=4.3cm]{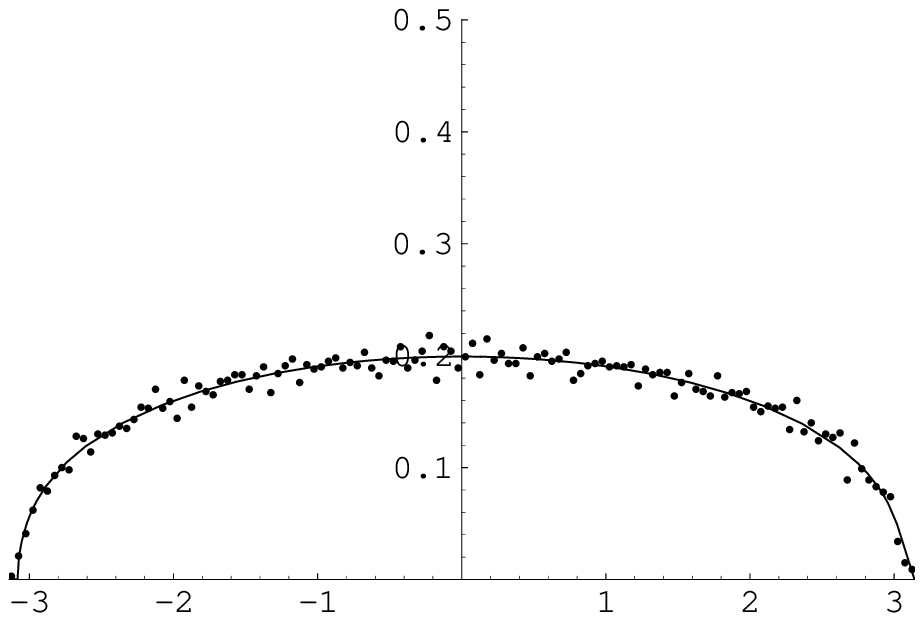}
\includegraphics[width=4.3cm]{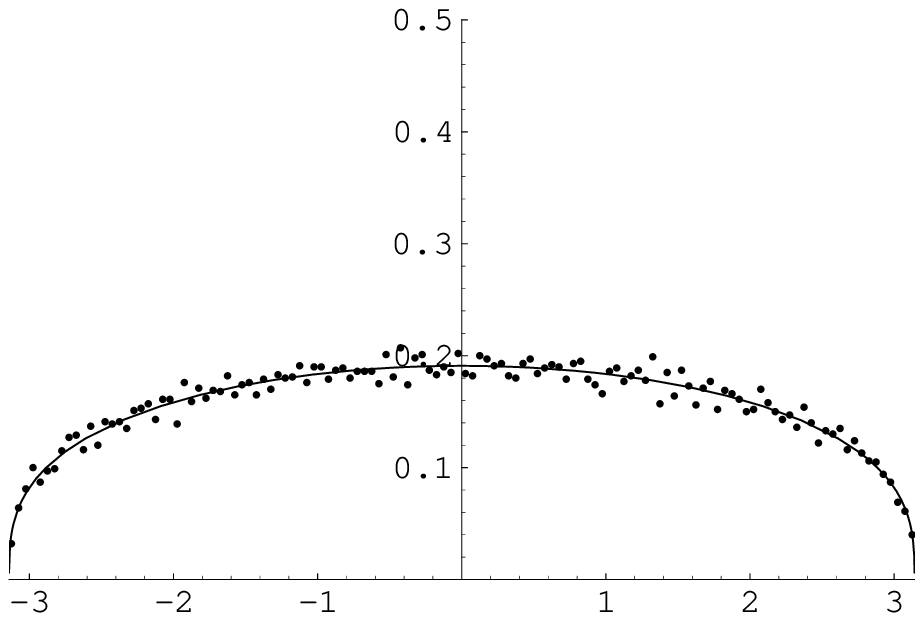}
\includegraphics[width=4.3cm]{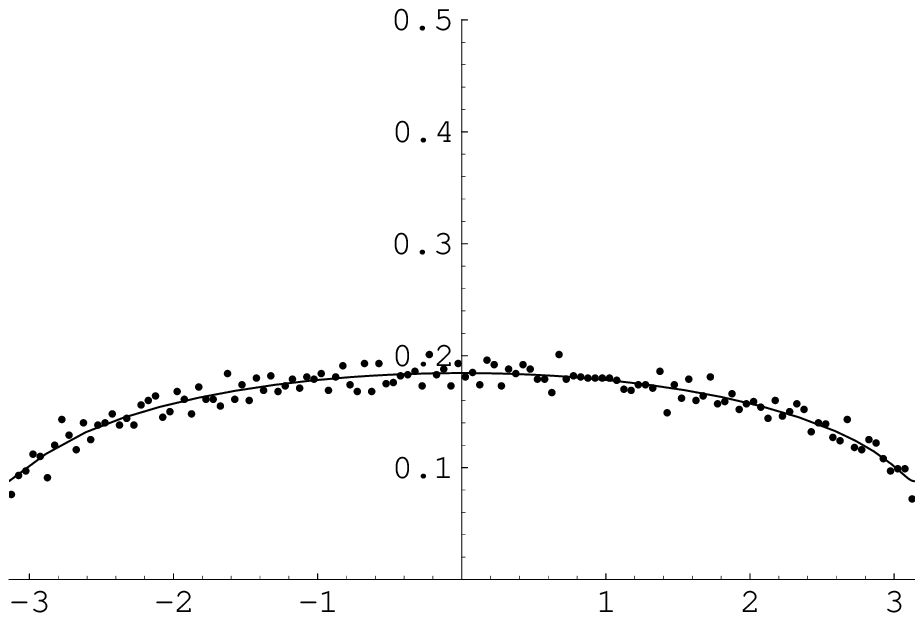}
\includegraphics[width=4.3cm]{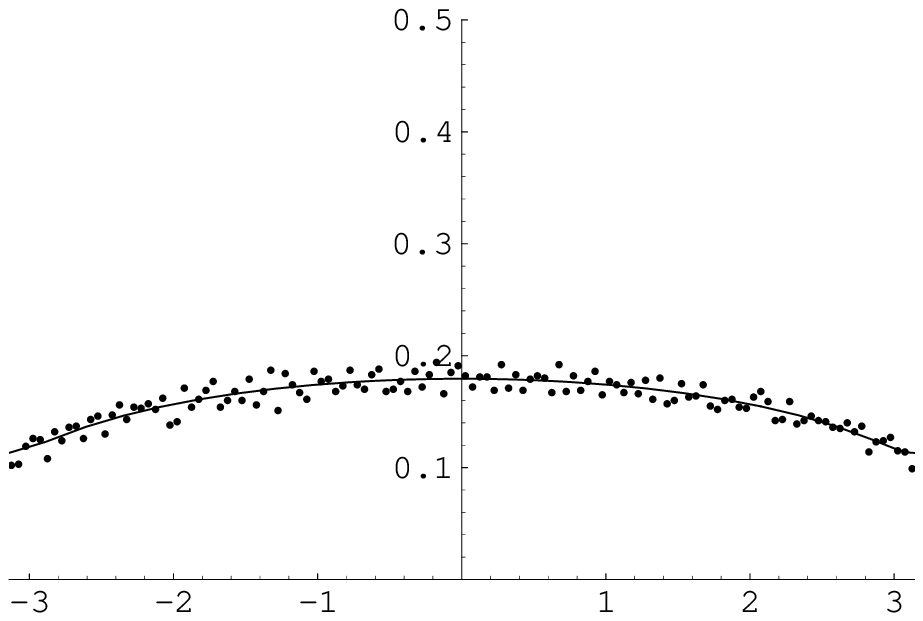}
\includegraphics[width=4.3cm]{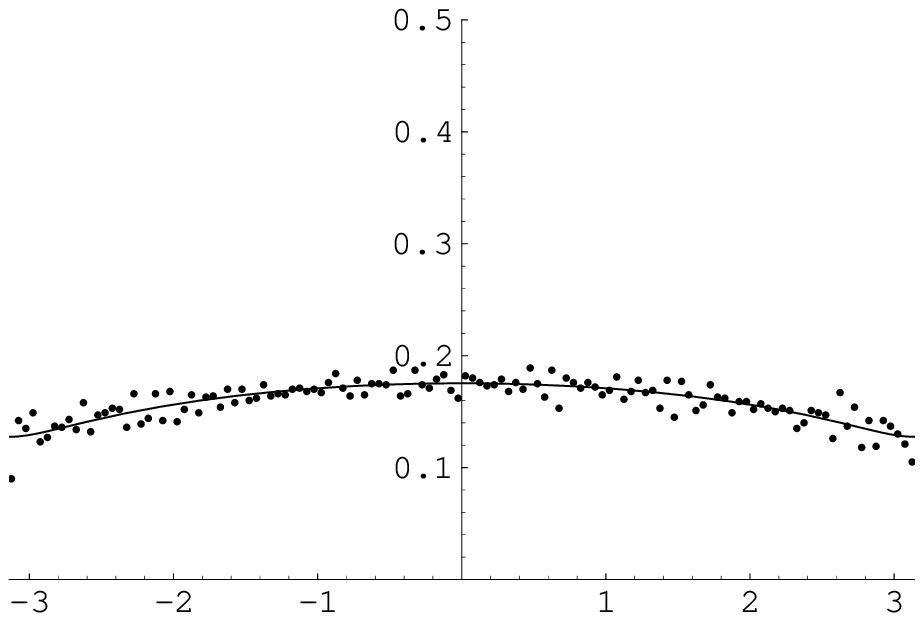}

 \caption{\label{evo} The time evolution of the spectral function
   $\rho(\theta,t)$ (the dots represent numerical simulation). The
   figures show
   $\rho(\theta,t)$ after a time $ t = 1, 250, 500$ up to 2750 (with
   $m_2 = 1/500$). The solid lines represent the eigenvalue density
   (\ref{e.rho}) obtained by the numerical solution of (\ref{e.wynik}).} 
\end{center}
\end{figure}

In this section we analyze the dynamical behavior of unitary matrix
diffusion. For small times $t$ the eigenvalues will be concentrated
only in a small neighbourhood of $\th=0$. 
For longer times the support of the eigenvalue density $\rho$ will expand and 
when some critical time $t_{c}$ is reached the eigenvalues will fill
the whole circle. But of course the eigenvalue density $\rho$
will be nonuniform. In fact we expect a critical behavior close to
$\th=\pi$ with nonstandard fractional eigenvalue spacing.
Only later for $t \rightarrow \infty$ the eigenvalues will become
uniformly spread over the whole unit circle. In fig. \ref{evo} we show
numerical results for the eigenvalue density obtained by generating
unitary matrices and compare it to the one extracted from
(\ref{e.wynik}) (see below).  

In this section we will quantitatively analyze this behavior.

\subsection*{The support of the eigenvalue distribution and the
  critical time~$t_c$}

Although one cannot find an analytical formula for the eigenvalue
density one can analytically find the edges of the eigenvalue
support. These occur when the Green's function has an infinite
derivative $\partial_z G=\infty$. Differentiating (\ref{e.wynik}) with
respect to $z$ gives
\be
1 = - \frac{\partial_{z} f}{f^{2}} \left( 1+  m_2 t f
  +m_2 t f^2 \right)  e^ { - t (f + \frac{1}{2}) m_2} \, .
\ee
So the end points are determined through the solutions of equation $1+
m_2 t f + m_2 t f^2 = 0$. Once we know $f$ we can reconstruct the
end-points $z$ using (\ref{e.wynik}). The result is
\eq
z_{edge}= \f{\sqrt{4-m_2 t}+i\sqrt{m_2 t}}{\sqrt{4-m_2 t}-i\sqrt{m_2
    t}} \cdot e^{\f{i}{2} \sqrt{m_2 t} \sqrt{4-m_2 t}}
\eqx
and its complex conjugate $z^*_{edge}$. 
When these two solution are equal ($z_{edge}=z^*_{edge}=-1$), the
eigenvalues will cover the whole circle.
This will happen for the critical time
\be
t_{c} = \frac{4}{m_2} \, .
\ee

\subsection*{The moments of $U$}

Another quantity which can be analytically calculated are the moments of
$U$. The coefficients of the auxiliary function $f$ around $z=\infty$ 
\be
\label{ans} 
f(z) = \sum_{k=1}^{\infty} \frac{a_{k}}{z^{k}} \, , 
\ee  
are indeed directly linked to the moments:
\be
a_{k} = \cor{\f{1}{N} \tr U^{k}} \, .
\ee
So inserting (\ref{ans}) into (\ref{e.wynik}) allows us to find the
moments iteratively. The expressions for the lowest ones are:
\begin{eqnarray}
\label{result1}
  \label{one} a_{1} &=& e^{-\frac{1}{2}  m_2\,t  } \\
  a_{2}&=& e^{- m_2\,t}\left(-1 + m_2\,t \right) \\
  a_{3}&=& \frac{1}{2} e^{-\frac{3\,m_2\,t}{2}}(2 - 6\,m_2\,t +
  3\,m_2^2\,t^2) \\ 
\label{result2}
  a_{4}&=& -\frac{1}{3} e^{- 2\,m_2\,t}\left( -3 + 18\,m_2\,t -
  24\,m_2^2\,t^2 + 8\,m_2^3\,t^3 \right) \\ 
  a_{5}&=& {\frac{1}{24} e^{- \frac{5\,m_2\,t}{2}}}(24 - 240\,m_2\,t +
  600\,m_2^2\,t^2 - 
         500\,m_2^3\,t^3 + 125\,m_2^4\,t^4) 
\end{eqnarray}

In fig. 2 we compare the formulae for the lowest 4 moments with
numerical simulations of the unitary matrix diffusion and find
complete agreement.

\begin{figure}[t!]
\begin{center}
\includegraphics[width=6.2cm]{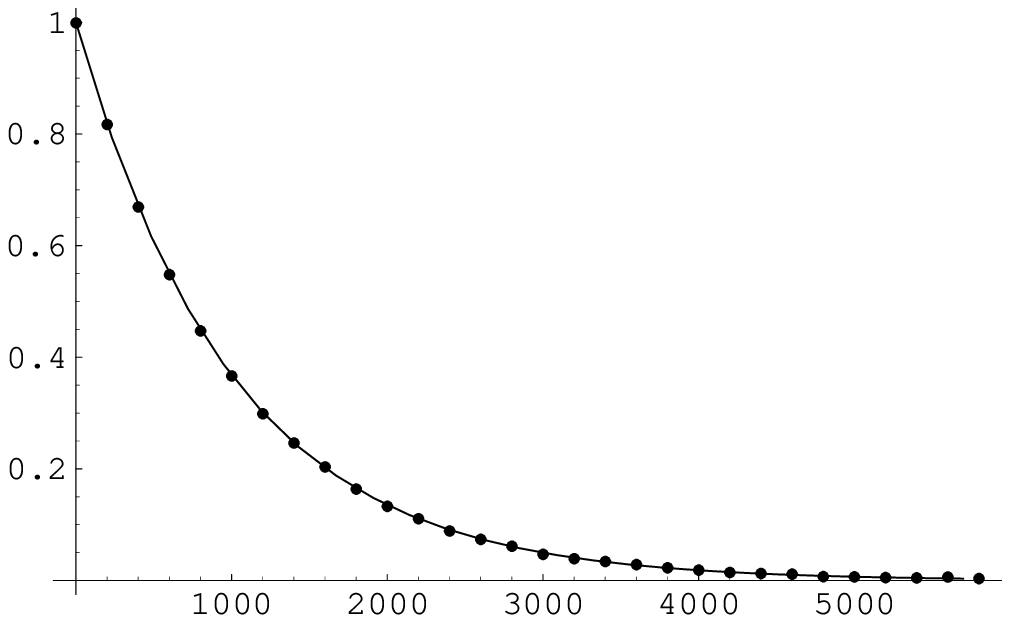}
\includegraphics[width=6.2cm]{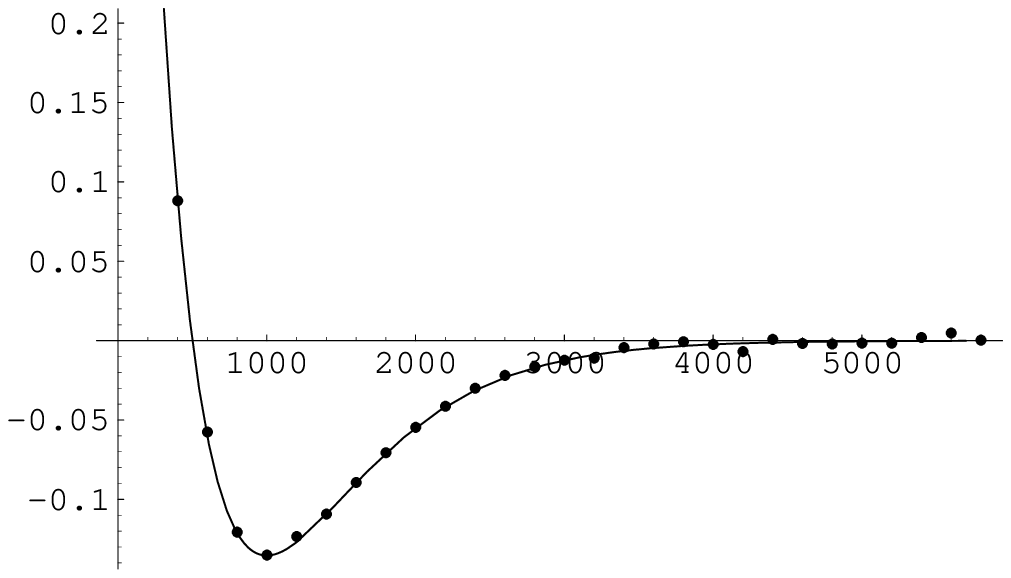}
\includegraphics[width=6.2cm]{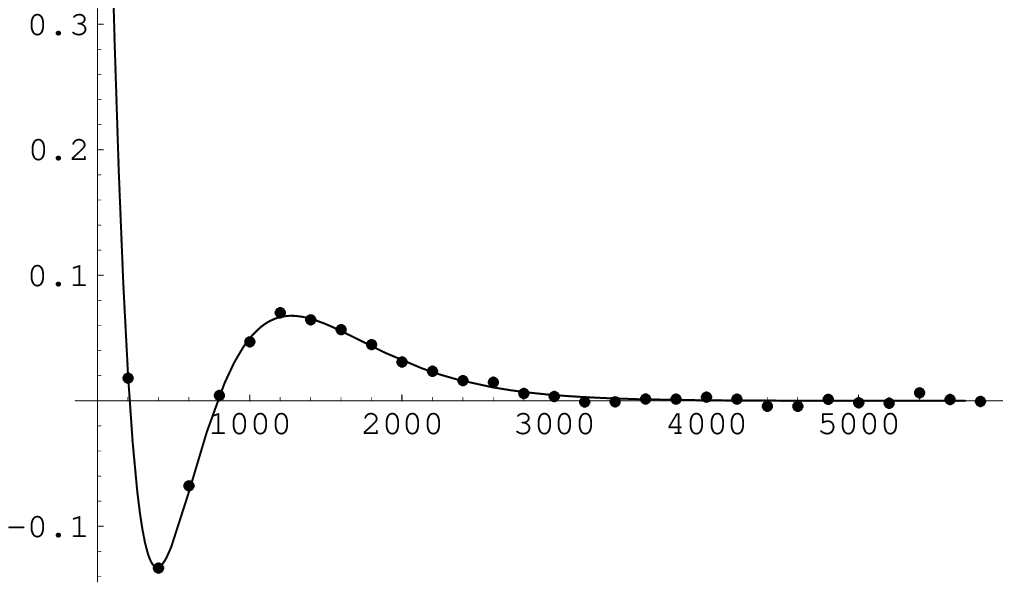}
\includegraphics[width=6.2cm]{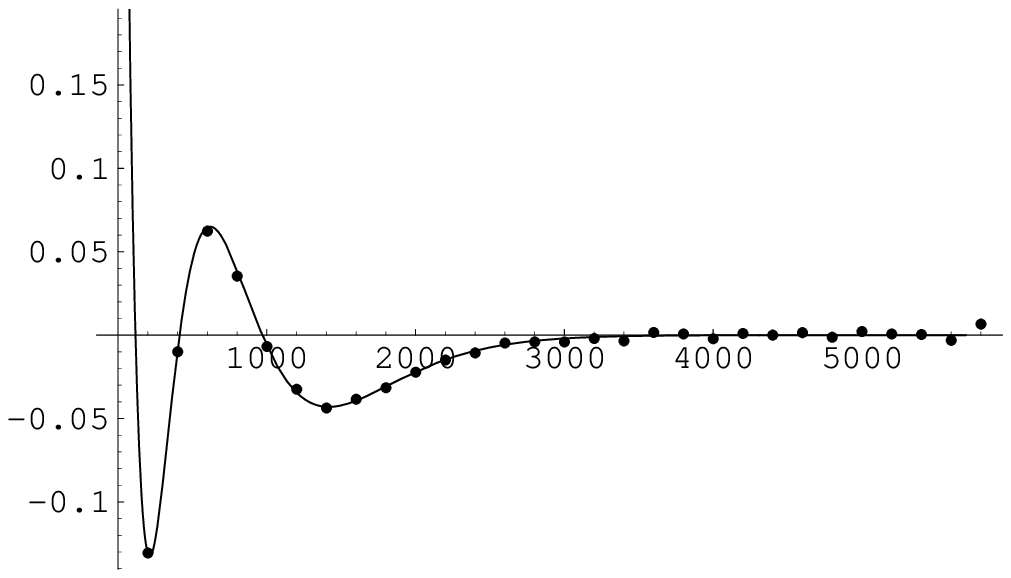}

\caption{The comparison of numerical simulations (dots)
with analytical results (lines), for $n=1,2,3,4$ and $m_2 = 1/500$.} 
\end{center}
\end{figure}

\subsection*{The eigenvalue density}

The equation (\ref{e.wynik}) allows us to directly reconstruct the
Green's function. However it is very simple to recover also the
eigenvalue density. This follows from the observation that the moments
of $U$ are just the Fourier coefficients of the eigenvalue density
$\rho(\th)$
\eq
\cor{\f{1}{N} \tr U^k} =\int_0^{2\pi} \rho(\th) e^{ik\th} \, .
\eqx
Using the relation of $f$ to moments derived earlier, and the symmetry
$\rho(\th)=\rho(-\th)$ one finds finally
\eq
\label{e.rho}
\rho(\th) = -\f{1}{2\pi} \mbox{\rm Re } \left(\f{1}{2}
  +f\right) \, .
\eqx
Equation (\ref{e.wynik}) may be easily solved numerically. In
fig. 1, we show the resulting eigenvalue density together with
numerical simulations for various times $t$.
 
\subsection*{Critical behavior at $t=t_c$ and level spacing}

At $t=t_c$ the edges of the eigenvalue support touch at
$z=-1$. Typically in such cases we expect new critical type of
behavior and nonstandard scaling of eigenvalue spacing with $N$. Let
us analyze now this behavior. Inserting $t = t_{c}$ and $f= - 1/2
+F$ to (\ref{e.wynik}) we obtain: 
\be
\label{assym}
z= \frac{F + 1/2}{F-1/2} \; e^{-4F} \; ,
\ee
To find the behavior close to $z= -1$ (equivalent to $\theta = \pi$) we
expand the left hand side of (\ref{assym}) in $F$ and put $z= -1 + i
y$ to get
\eq
-1 + i y \approx -1 - \frac{16 F^{3}}{3} \; .
\eqx
Using the relation between $f$ and the eigenvalue density
(\ref{e.rho}) we thus find the behavior close to $\th=\pi$:
\eq
\rho(\th) \sim  \left\{ \f{1}{2\pi} \left(\f{3}{16}\right)^{\f{1}{3}}
\cos \f{\pi}{6} \right\} \cdot  |\th-\pi|^{\f{1}{3}} \, .
\eqx

Such behavior of the eigenvalue density leads to nonstandard
eigenvalue spacing and signifies the appearance of new universal
regime on the scale of eigenvalue spacing (analogous to Airy
universality and $1/N^{2/3}$ spacing on the edges of the eigenvalue
distribution of a generic hermitian random matrix
\cite{AIRY,BI1,DEIFT,PASTUR} in contrast to the standard $1/N$ spacing
in the classical Wigner-Dyson regime \cite{WD}). 

In our case, the number of eigenvalues between $\pi$ and $\Lambda$ is
approximately equal to $n \sim N (\Lambda-\pi)^{4/3}$. Reexpressing
$\Lambda$ in terms of $n$ shows that the eigenvalue spacing in the
vicinity of $\th=\pi$ scales like $1/N^{3/4}$. A similar scaling
appeared in a certain class of chiral random matrix models at finite
temperature \cite{USCRITICAL}. It would be interesting to compare
these regimes and/or try to apply the methods of \cite{MULTICRIT} to
the case at hand. We leave this problem for further investigation.  

\section{Discussion}

In this paper we considered multiplicative unitary matrix diffusion
generated by random hermitian matrices. We found the eigenvalue
density as a function of evolution time in the large $N$ limit using
$S$-transform methods. The eigenvalue distribution turns out to be
{\em universal} and depends only on the second moment of the random
hermitian matrix which generates the diffusion process. 

We found that at a critical time of evolution $t=t_c$ the eigenvalues
start to fill the whole unit circle, and close to $\th=\pi$ a
nonstandard eigenvalue spacing $\sim 1/N^{3/4}$ sets in which signifies
the appearance of a new critical regime. 

There are various further issues that one could investigate. Firstly,
relaxing the assumption of the existence of the second moment might
lead to defining anomalous diffusion processes. Secondly it would be
interesting to study microscopic properties of these unitary matrices,
however in order to do that new methods have to be developed. Thirdly
a more detailed investigation of the critical behavior at $t=t_c$
close to $\th=\pi$ would be interesting and last but not least the
application of these results to some physical situations. 

\bigskip

\noindent{\bf Acknowledgments.} We would like to thank Maciej A. Nowak
for numerous discussions. RJ would like to thank 
the Niels Bohr Institute for hospitality while this work was
being completed. This work was supported in part by KBN grants~2P03B09622
(2002-2004), 2P03B08225 (2003-2006). RJ was supported by ``MaPhySto'', 
Network in Mathematical Physics and Stochastics financed by the Danish
National Research Foundation.

\end{document}